\DocumentMetadata{}
\documentclass[acmsmall]{acmart}

\AtBeginDocument{%
  \providecommand\BibTeX{{%
    \normalfont B\kern-0.5em{\scshape i\kern-0.25em b}\kern-0.8em\TeX}}}

\setcopyright{cc}
\setcctype{by}
\acmJournal{PACMHCI}
\acmYear{2025} \acmVolume{9} \acmNumber{4} \acmArticle{EICS007} \acmMonth{6} \acmPrice{}\acmDOI{10.1145/3733053}

\usepackage{multirow}
\usepackage{makecell}
\usepackage{tabularx}

\usepackage{multirow}
\usepackage{graphicx}

\usepackage{tabularray}

\usepackage{caption}
\usepackage{subcaption}

\usepackage[capitalize, noabbrev]{cleveref}

\usepackage{layouts}

\usepackage{listings}

\UseTblrLibrary{booktabs}

\definecolor{applegreen}{rgb}{0.55, 0.71, 0.0}
 	\definecolor{darkelectricblue}{rgb}{0.33, 0.41, 0.47}
  \definecolor{coralpink}{rgb}{0.97, 0.51, 0.47}
  \definecolor{darkgray}{rgb}{0.66, 0.66, 0.66}

\newcommand{\sal}[0]{\textit{StudyAlign}}

\newcommand{\controlpanel}[0]{Control Panel}

\newcommand{\studyfrontend}[0]{Study Frontend}

\newcommand{\backend}[0]{Backend}

\newcommand{\library}[0]{Library}

\begin{teaserfigure}
   \includegraphics[width=1\textwidth]{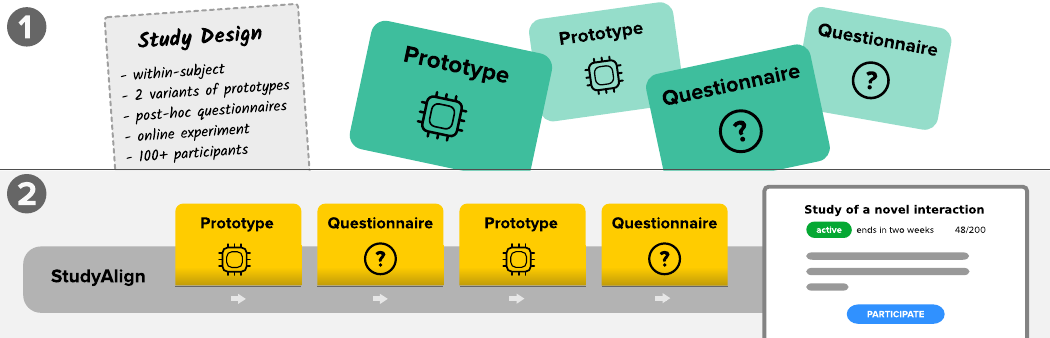}
   \caption{From 1) a sketched study design to 2) a complete, structured web-based experiment with interactive prototypes: We introduce \sal, a software system to conduct web-based experiments with functional interactive prototypes. \sal{} offers four software parts for creating, organizing, and conducting online studies: a \backend{} to create studies and data logging, a \library{} to integrate prototypes, a \studyfrontend{} that guides participants, and an administrative \controlpanel.}
    \Description{This figure illustrates that \sal{}streamlines web-based experiments. The top half shows a sheet of paper with notes of a sketched study design. On the right of these notes are four cards visible: two prototype cards and two questionnaire cards. These cards visualize already prepared prototypes and questionnaires by a research team, yet they are not part of an experiment. The bottom half visualizes that \sal{} works as a foundation for web-based experiments and puts these prototypes and questionnaires in order. The four cards are bound together in one row from left to right and result in a participant's web-view, which is visible in the bottom right of the figure.}
    \label{fig:teaser}
\end{teaserfigure}

\begin{document}

\title{StudyAlign: A Software System for Conducting Web-Based User Studies with Functional Interactive Prototypes}

\renewcommand{\shorttitle}{StudyAlign: A Software System for Conducting Web-Based User Studies}

\author{Florian Lehmann}
\orcid{0000-0003-0201-867X}
\email{florian.lehmann@uni-bayreuth.de}
\affiliation{%
  \institution{University of Bayreuth}
  \city{Bayreuth}
  \country{Germany}
}

\author{Daniel Buschek}
\orcid{0000-0002-0013-715X}
\email{daniel.buschek@uni-bayreuth.de}
\affiliation{%
  \institution{University of Bayreuth}
  \city{Bayreuth}
  \country{Germany}
}

\renewcommand{\shortauthors}{Lehmann and Buschek}

\begin{abstract}
Interactive systems are commonly prototyped as web applications. This approach enables studies with functional prototypes on a large scale. However, setting up these studies can be complex due to implementing experiment procedures, integrating questionnaires, and data logging. To enable such user studies, we developed the software system \textit{\sal} which offers: 1) a frontend for participants, 2) an admin panel to manage studies, 3) the possibility to integrate questionnaires, 4) a JavaScript library to integrate data logging into prototypes, and 5) a backend server for persisting log data, and serving logical functions via an API to the different parts of the system. With our system, researchers can set up web-based experiments and focus on the design and development of interactions and prototypes. Furthermore, our systematic approach facilitates the replication of studies and reduces the required effort to execute web-based user studies. We conclude with reflections on using \sal{} for conducting HCI studies online.
\end{abstract}



\maketitle

\section{Introduction}

Imagine you have designed and prototyped novel interaction methods with a functional web prototype. Now, you want to investigate these interaction methods in a user study to collect quantitative and qualitative data. Since your prototypes are built with web technologies, you consider running an online study. However, setting up an online study that combines prototypes (logging quantitative data) with questionnaires (collecting qualitative data), implementing study designs (counter-balancing of conditions and controlling procedures), and conducting the study adds additional effort. Concretely, this effort arises in large parts from building additional software, for example, for data logging backends, implementing the counter-balancing dynamically or through hard coding. Furthermore, these efforts surrounding an experiment tend to re-occur for each study.

In HCI research, user interfaces and interactions are potentially built for the web and examined through online studies (e.g. \cite{lee_coauthor_2022, robertson_i_2021, huang_scones_2020}). A reason to build web-based prototypes is the availability of frontend libraries such as React\footnote{\url{https://react.dev/}} and Bootstrap\footnote{\url{https://getbootstrap.com/}} that lower barriers for prototyping since many examples and tutorials exist and they are reliable to develop with. Furthermore, researchers have the possibility to run online studies and scale their studies easily.
In particular, work on human-AI interaction has recently been based on web prototyping (e.g. \cite{masson_directgpt_2024, vaithilingam_dynavis_2024, benharrak_writer-defined_2024, dang_choice_2023, gero_sparks_2022}). These studies rely on web technologies since they require AI services that are mostly available via the web, for example, platforms like Hugging Face\footnote{\url{https://huggingface.co/}} distribute machine learning (ML) models for download, with eased implementation efforts, and instructions for developers. A common approach is to deploy ML models on web servers and make them available via programmable interfaces (APIs). Also, commercial services of such APIs exist. HCI researchers use such APIs to build a model's capability into web application prototypes that enable novel interaction methods or UIs. The process of informing, designing, developing, and analyzing such prototypes typically requires multiple studies to be carried out (e.g. \cite{schmitt_characterchat_2021, todi_conversations_2021, han_textlets_2020}). 

Motivated by applied methods from such related work and our own experience in conducting online research, we introduce \sal, a software system for conducting web-based experiments. With \sal{}, we follow the philosophy to reduce the previously mentioned efforts surrounding setting up web-based experiments. This way, HCI researchers can focus on prototyping UIs and interaction methods. 

Our system is built from four distinct components:

\begin{itemize}
\item The \backend: Is the core of the system. It offers database storage for interaction logging and an API to provide important methods to all other components. The \backend{} ties together conditions, questionnaires, and text pages and offers counterbalancing.

\item The \controlpanel: Is a browser app that offers a UI for administrative purposes. Researchers use the \controlpanel{} in a web browser to configure and organize their experiments. A crucial feature is the import and export of studies. This feature allows for sharing and replicating experiments.

\item The \studyfrontend: Is used by participants to take part in a study. The \studyfrontend{} guides participants step-by-step through the experiment and increases control for researchers (e.g. by preventing participants from skipping steps).

\item The \library: Is integrated into prototypes via JavaScript. Researchers include the \library{} into their prototypes to connect them to the \backend, and log interactions. The \library{} is an abstraction layer and makes logging easy, for example, logging native browser events.
\end{itemize}

At the time of writing, the system has been applied in more than ten HCI studies. Some of these studies have already been published (e.g. \cite{draxler_ai_2023, dang_beyond_2022, dang_choice_2023, benharrak_writer-defined_2024, zindulka_content-driven_2025, zindulka_exploring_2025}). Based on this hands-on experience and collaborations with other researchers, we iterated on the features of our system. Furthermore, these deployments demonstrate the system's success in supporting researchers in conducting online experiments.

With \sal, we contribute a software system for conducting web-based HCI research. Our system makes setting up such experiments more convenient, allows for interaction logging, and makes sharing and replication of experiments possible directly as files. In this paper, we give detailed insights into how the distinct building blocks of \sal{} work internally and connect to provide the system as a whole. Furthermore, we sketch potential use cases and reflect on previous studies with our system to demonstrate its usability. Finally, we discuss the possibilities offered by \sal{} and the methodological aspects such as replication and study sharing. We make the project available as an open-source project: \url{https://studyalign.com}.

\section{How to set up an experiment with \sal}

\begin{figure*}[t]
    \begin{center}
    \includegraphics[width=1\textwidth]{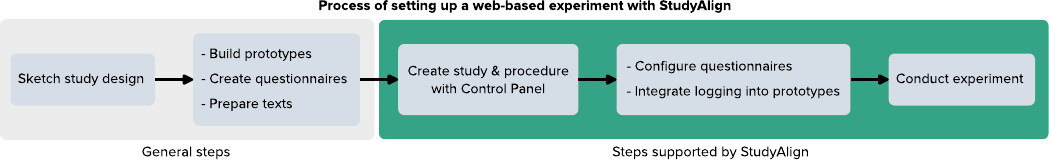}
    \caption{This figure visualizes the process of researchers for setting up web-based experiments with \sal. The first two blocks in this process cover sketching the study design, designing and implementing prototypes, creating questionnaires in tools such as Qualtrics, and preparing text documents, such as task briefings. The latter three blocks are then supported by \sal. Researchers use the \controlpanel{} for creating the study and procedure. Instructions support the researchers integrating questionnaires into the procedure, and the \library{} allows researchers to enable interaction logging in the prototypes. Finally, researchers conduct the experiment by sharing a study link with participants that displays a \studyfrontend{} to guide them through the experiment.}
    \Description{This figure shows a flow chart visualizing the general process of implementing a study with \sal. It consists of five steps from left to right. The first two steps are general steps. The last three steps are supported by \sal. The first two steps visualize ``sketch a study design'' followed by a step combining the sub-steps: ``build prototypes'', ``create questionnaires'', and ``prepare texts''. The third step shows ``create study and procedure with \controlpanel'', followed by a step combining the two sub-steps ``configure questionnaires'' and ``integrate library into prototypes'', and the final step shows ``conduct experiment''.}
    \label{fig:walk-through}
    \end{center}
\end{figure*}

In this section, we illustrate with a basic example how to set up an experiment with \sal. We visualized the process for setting up an experiment as a flow chart in \cref{fig:walk-through}. Our example study has a mixed-methods within-subject design. The requirements for this study are as follows: The goal of the study is to compare two different versions of an interaction. Quantitative data should be recorded while participants are using the prototype, and questionnaires should assess the usability and collect feedback from participants after using each prototype. Another questionnaire assesses socio-demographic data and more general feedback. A study design for this kind of experiment is depicted in \cref{fig:use-cases}~A, and is a typical design for HCI studies.

Before setting up the experiment with \sal, the research team implements the prototypes with their preferred web-technology stack and creates the questionnaires with a survey system of their choice (e.g. Qualtrics). Furthermore, they need to prepare texts for introducing the study to participants, the consent information, and the task briefings.

\subsection{Creating a new study}
To set up the study, the researchers use the \controlpanel{}, which guides them through four steps. These steps are depicted in \cref{fig:screenshot-procedure1} and \ref{fig:screenshot-procedure2}. In the first step, the setup wizard asks for general information such as the study's title, dates, the planned number of participants, an introduction text, and a consent text. 

In the second step, an interactive view supports the researchers in setting up the study. Concretely, this view allows the researchers to create text elements for the task briefings, condition elements for the prototype versions, and questionnaire elements for the different questionnaires. To add prototypes and questionnaires to the procedures, the researchers have to specify URLs in the elements (e.g. the URL to their hosted prototype web app). Moreover, block elements enable the researchers to group several elements, such as grouping the task briefing, condition, and questionnaire for each prototype version. They can then select these blocks to apply counterbalancing.

The third step gives instructions on integrating the questionnaires with \sal. The researchers need to complete these instructions in the survey tools, for example, setting up questionnaire variables that can retrieve data from the URL and save them to the database. For example, in Qualtrics, this is done by following five easy steps.

The researchers can check their setup in the fourth and last step before finalizing the study. Without \sal{} and its \controlpanel{}, the research team would have to technically implement the procedure, supposedly hard-coded by chaining the prototypes, questionnaires, and briefings through chaining links from one study component to another, and take care of realizing the counterbalancing. The more complex a study design gets, the more tedious such manual efforts become, for example, with three prototype variants (i.e. three conditions), or even more complex study designs.

\subsection{Integrating interaction logging}
After the study has been set up, the researchers need to include the \library{} into their prototypes and put logging statements at the desired parts in the code, for example, to capture clicks on a certain button. Without our system, the researchers would have to manually create a backend for persisting log data, and implement the requests for calling the backend methods.

\subsection{Running the study}
The system generates a study link to be shared with the participants. If participants open this link, they see the \studyfrontend{} that guides them through the complete study with a toolbar at the bottom of the screen. Without \sal{}, the researchers would have to think about a way to identify the participants across the procedure, for example, passing an ID through all steps via URLs and implementing when to show the link to the next step, for example, by taking care of a condition's progress. Furthermore, they would have to export the data from their logging backend manually, for example, by writing database queries or connecting to the server to download the logfiles.

To scale a running study, researchers can edit the study and increase the number of participants in the \controlpanel. At any time, the study scheme and log data can be exported from the system.

\subsection{Summary}
Overall, this walk-through shows that \sal{} is connecting many steps in the process of setting up and managing studies (\cref{fig:walk-through}), by offering convenient UIs (\cref{fig:screenshot-procedure1} and \cref{fig:screenshot-procedure2}), and robust methods to research teams.

\section{Related work}

To inform the design and development of our system, we screened related work on interactive AI prototypes and their evaluation (e.g. \cite{gero_sparks_2022, lee_coauthor_2022, zhang_method_2021}). Furthermore, we looked at web-based research and field experiments from a methodological point of view, as well as frameworks and tools for conducting web-based research. In the following paragraphs, we discuss each of these aspects in more detail and how they influenced \sal.

\subsection{Web-based (field) experiments}

The applicability of laboratory and field research in the domain of HCI has been discussed by the research community for decades \cite{wolf_role_1989}. Depending on what researchers investigate, one method can be more beneficial than the other \cite{reis_field_2014}. A main difference is the level of control (e.g. control of variables, assignment of participants) and validity these methods offer. Following \citet{shadish_experimental_2002}, validity can at least be further separated into internal validity (causal relationship between manipulated and measured variables) and external validity (generalization of cause-effect relationship). Laboratory studies are commonly considered to offer a high level of control and high internal validity, yet a low external validity. In comparison, field research is considered to offer a decreased level of control and decreased internal validity but an increased external validity. The combination of both methodological approaches results in field experiments. With field experiments, researchers have some control over the experiment while balancing internal and external validity. If participants cannot be assigned randomly, a field experiment is called a quasi-experiment \cite{shadish_experimental_2002}. For the sake of consistency, we stick to  ``web-based experiment'' as a term for ``field experiment'' throughout this paper. %

Such web-based experiments can be conducted online. In other research domains, for example, in psychology, standards for internet-based experimenting were proposed \cite{reips_standards_2002}, and plenty of software frameworks exist (e.g. \cite{henninger_labjs_2019, de_leeuw_jspsych_2015, mathot_opensesame_2012}). For stimuli-based experiments, comparability of reaction times in lab- and web-based setups was confirmed several times \cite{hilbig_reaction_2016}. In a recent survey by \citet{reips_web-based_2021}, web-based psychology research has been reviewed, for example, they discuss web surveys and questionnaire research, web-based tests, and web experiments. 

Similar to these experiments from psychology, web-based experiments are a promising approach for HCI research. However, online experiments require the development of technical platforms and tools \cite{isomaki_field_2009}. In particular, research at the intersection of HCI and AI found the general process of prototyping AI applications challenging \cite{dove_ux_2017, yang_re-examining_2020} but sees a potential for fast prototyping with web-based ML tools \cite{li_ml_2021}. The main challenges in prototyping interactive AI are rooted in the technical complexity \cite{dove_ux_2017} and uncertainty \cite{yang_re-examining_2020}. Compared to surveys, interactive AI prototypes are often complex to prototype. Moreover, conducting experiments adds further complexity, for example, maintaining the experiment procedure, realizing a systematic counterbalancing of the order of tasks, and logging interactions. 

While our system aims to enable experiments on human-AI interaction, it is not limited to that and supports any web-based prototype. Overall, we introduce a technical basis for conducting web-based experiments with a focus on functional interactive prototypes.

In the following, we examine papers on web prototypes to motivate the requirements of \sal{} and distinguish our system from existing frameworks from psychology, behavioral research, and cognitive sciences.

\subsection{Interactive AI prototypes are evaluated on the web}\label{subsec:ai-prototypes-are-evaluated-on-the-web}

To inform the design and development of our software system, we screened HCI research contributions that investigate interactions with web-based prototypes. In particular, we focused on research at the intersection of HCI and AI, as these current research efforts rely on web-based prototypes to investigate novel interactions, UIs, and applications. We examined these papers from a methodological point of view. We compared these studies and our own experiences to identify recurring similarities. These similarities defined the requirements and the scope of \sal. 

Without functional prototypes, yet with web experiments, related work investigated the effects of AI-mediated communication \cite{jakesch_ai-mediated_2019} and socially problematic AI-generated email reply suggestions \cite{robertson_i_2021}. Web experiments with functional AI prototypes studied text interaction \cite{schmitt_characterchat_2021, lee_coauthor_2022, todi_conversations_2021, arnold_sentiment_2018, gero_sparks_2022, lehmann_suggestion_2022, chung_talebrush_2022, han_textlets_2020, buschek_impact_2021}, image interaction \cite{dang_ganslider_2022, weber_draw_2020, zhang_method_2021, masson_directgpt_2024, suh_luminate_2024}, visualization editing \cite{vaithilingam_dynavis_2024}, creative applications with sound \cite{louie_novice-ai_2020}, drawing with language input \cite{huang_scones_2020}, and the evaluation of a tool to prototype the design of AI applications \cite{subramonyam_protoai_2021}. Most of these studies consist of multiple parts, such as subsequent studies, for example, to inform and iterate on the design of a prototype, and its evaluation \cite{jakesch_ai-mediated_2019, schmitt_characterchat_2021, todi_conversations_2021, zhang_method_2021, gero_sparks_2022, han_textlets_2020, buschek_impact_2021}. Data has been collected in all studies, including qualitative data through questionnaires and interviews, and quantitative data through data logging, oftentimes both. For recruiting participants, the researchers relied on platforms such as Amazon Mechanical Turk or Prolific \cite{lee_coauthor_2022, robertson_i_2021, huang_scones_2020}. While related work reported on experiment procedures, the authors were somewhat unclear on how they technically implemented it (e.g. \cite{buschek_impact_2021, dang_ganslider_2022, gero_sparks_2022, lehmann_suggestion_2022, han_textlets_2020, louie_novice-ai_2020, zhang_method_2021}), and randomized condition order instead of applying systematic counterbalancing (e.g. with a Latin Square) \cite{arnold_sentiment_2018, yuan_wordcraft_2022, weber_draw_2020}.

These aspects, namely conducting studies that consist of multiple experiments, interaction logging, online recruitment of participants, and implementation of experiment procedures, defined the core requirements of \sal. As implemented, \sal{} enables researchers to set up the procedures of all of these studies (and more). In particular, studies can be configured and managed through an administrative \controlpanel{}, and the experiment logic is provided by the \backend. This enables researchers to run subsequent studies or multiple studies at the same time conveniently (e.g. as in \cite{schmitt_characterchat_2021, todi_conversations_2021, han_textlets_2020}) without re-implementing the experiment logic. Data logging is integrated into the system, as required by all of the mentioned studies, and experiments can be shared on platforms for recruiting participants (e.g. as in \cite{lee_coauthor_2022, robertson_i_2021, huang_scones_2020}). Furthermore, our software system makes web-based experiments more robust by controlling task order with counterbalancing (e.g. as required in \cite{buschek_impact_2021, dang_ganslider_2022, gero_sparks_2022, arnold_sentiment_2018, yuan_wordcraft_2022}). Finally, study designs can be exported and shared easily along with publications.

For the technical implementation, we analyzed similar frameworks and considered their implementation approaches and how we could adapt them for studies with functional interactive prototypes. However, none of these frameworks offered a basis we could extend from, as the next subsection shows.

\subsection{Frameworks and tools}

Work by \citet{ledo_evaluation_2018} offers a good overview of existing software frameworks and toolkits. Furthermore, the authors identified four common evaluation strategies for such toolkits. Their paper was the initial starting point for our literature research on frameworks, and we also applied two of their identified strategies to evaluate our framework. For our research, we focused on software frameworks that were close to the requirements we derived from papers that evaluated interactive AI prototypes. In particular, we considered experiment frameworks and looked at commercial products and services.

\subsubsection{Research frameworks}

Most of the software frameworks for conducting web-based experiments stem from research on psychology, behavioral research, and cognitive sciences. We categorized these frameworks primarily by their purpose.

Natural field studies on mobile devices are enabled by frameworks such as \textit{AWARE}~\cite{ferreira_aware_2015} and \textit{PhoneStudy}~\cite{stachl_predicting_2020}. As a more general tool for experiment design, \textit{TouchStone2} allows researchers to interactively generate and examine trade-offs in experiment designs \cite{eiselmayer_touchstone2_2019}.

Other frameworks focused on backend implementations, experiment application building, and source development kits and libraries:  
For backend implementations, we found \textit{Pushkin}, a backend aiming for large-scale user studies \cite{hartshorne_thousand_2019}, and \textit{JATOS}, for setting up and managing studies \cite{lange_just_2015}. Another minimalistic backend implementation combined with an experimenter dashboard is offered by \textit{ReActLab}, a framework for programming web-based experiments based on other libraries \cite{balestrucci_reactlab_2022}.

The popular Python source development kit \textit{PsychoPy} \cite{peirce_psychopypsychophysics_2007} and a JavaScript library for creating behavioral experiments \cite{de_leeuw_jspsych_2015}, enable researchers to implement a wide range of experiment application for behavioral sciences, for example, stimuli-based experiments with reaction time tasks.

In general, the larger part of the literature presents frameworks that allow for building interactive experiment applications. These are \textit{Empirica}, a framework to build interactive multiplayer experiments \cite{almaatouq_empirica_2021}, \textit{UILab}, a framework for creating and conducting studies comparing visual designs \cite{burny_uilab_2021}, \textit{SimplePhy}, a builder for perception experiments \cite{lago_simplephy_2021}, \textit{OpenSesame}, a graphical experiment builder for social sciences \cite{mathot_opensesame_2012}, and \textit{lab.js}, a modern online study builder for the behavioral and cognitive sciences \cite{henninger_labjs_2019}. An experiment builder specifically for HCI-related elicitation studies is \textit{Crowdlicit} \cite{ali_crowdlicit_2019}. 

\subsubsection{Commercial services}

In addition to the research frameworks, we also identified commercial online services such as the prototyping tools \textit{Maze}\footnote{\url{https://maze.co}} and \textit{ProtoPie}\footnote{\url{https://www.protopie.io}}. Both tools offer the setup and management of experiments with prototypes in the design stage. In particular, these tools allow designers to add simple interactivity to their visual designs and combine the prototypes with questionnaires in a procedure. While this approach enables designers to validate design decisions in an early stage of product design, it remains a challenge to prototype interactive AI applications \cite{dove_ux_2017}, for example, because of the uncertainty and the output complexity of AI's capabilities \cite{yang_re-examining_2020}. In this regard, the research prototyping tool \textit{ProtoAI} was introduced to include AI models in the design stage \cite{subramonyam_protoai_2021}. The authors of \textit{ProtoAI} conclude, however, that design prototypes may be unable to cover each case that could appear when users interact with AI. They suggest investigating interactive prototypes, sharing them with end-users, and including logging capabilities. In the same line, we argue that functional implementation of these prototypes is required to design, develop, and understand the interaction properly. Furthermore, more extensive studies may be required to understand complex interactions that go beyond design prototypes. 

\sal{} supports researchers with conducting such extensive studies and interactive prototypes beyond the design stage. In the next paragraph, we highlight this concretely and distinguish the scope of our framework from prior work.

\subsubsection{The scope of \sal{} as a framework for web-based HCI experiments}

From a technical perspective, \sal{} offers functionalities surrounding web-based HCI experiments. These functionalities are different from already existing systems. %

Since functional prototypes are more complex, particularly when they integrate AI capabilities, we do not offer a graphical experiment-building tool, such as \cite{lago_simplephy_2021, mathot_opensesame_2012,henninger_labjs_2019}. 
\sal{} offers a library to integrate prototypes into the system and enables interaction logging. It does not offer methods to implement applications specifically, such as \textit{PsychoPy} or \textit{jsPsych} \cite{peirce_psychopypsychophysics_2007, de_leeuw_jspsych_2015}. Creating interactions and building prototypes remains a task of the research teams. However, \sal{} makes it easier for HCI researchers to conduct web-based research with such prototypes, so they can focus on the design, implementation, and analysis of interactions. %

\begin{figure*}[t]
    \begin{center}
    \includegraphics[width=0.75\textwidth]{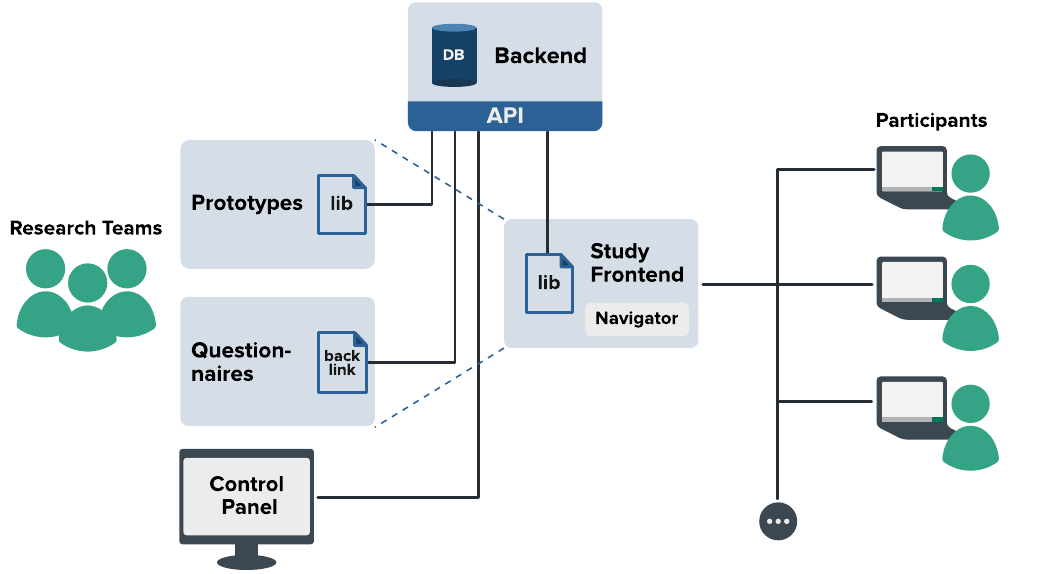}
    \caption{Overview of the software system \sal. The main components are the \backend, \controlpanel, \studyfrontend, and \library. The \backend{} is the core of the system since it holds the logic and persists data in a database. The \controlpanel{} is used by researchers to set up and manage studies. The \studyfrontend{} controls the experiment procedure (see \Cref{fig:navigator} about its Navigator for details) and is served to the participants' devices. It embeds prototypes and questionnaires and displays them to participants. Questionnaires use a callback function to work with the API. The \library{} is the counterpart to the \backend's API and is used in the prototype for interaction logging and in the \studyfrontend{} for controlling the procedure.}
    \Description{This figure shows an overview of the software system \sal. Each software component is visualized as a card and connected through lines to visualize relationships. In the center top is the \backend{} card with a database icon and a border at the bottom to visualize the API. The API bar connects through lines to the other components. On the left of the figure is an icon with three abstract people that visualize the research team. On the right of this icon are the two cards for prototypes and questionnaires. Within the prototype card is a file icon connected to the backend API, the file icon visualizes the \library. The questionnaire card is connected to the API through a file icon that visualizes a backlink. Below these cards is a screen icon visualizing the \controlpanel{} that is also connected to the backend. In the center of the figure is the card that visualizes the \studyfrontend{}, it also has a file icon connected to the API, and another area visualizing the Navigator. Dashed lines from the prototypes and questionnaires visualize a funnel into this \studyfrontend{} card. Lines connect on the right of the \studyfrontend{} card to the participants visualized through laptop icons with a person in front of it.}
    \label{fig:system-overview}
    \end{center}
\end{figure*}

\section{The \sal{} system}
\sal{} is a completely web-based software system for conducting HCI research. It is not limited to running studies online. It is also possible to utilize it for laboratory setups. In this section, we will state our motivation, give a high-level overview of our system, and provide detailed insights into its software parts. A feature overview can be found in \cref{tab:framework-features}.

\begin{table}[hb]
\caption{\sal{} comes with a broad set of features. We highlight the most important features on a system level.}
\Description{Description}
\label{tab:framework-features}
\begin{tblr}{lX}
\toprule
\textbf{Feature}           & \textbf{Description}   \\ \hline
UI for creating web-based studies & The \controlpanel{} supports setting up and managing web-based studies and allows for various study designs. Procedures can be edited with drag and drop, and parts can be selected easily for counter-balancing. \\
Library for integrating prototypes & The \library{} can be included in existing prototypes and enables the integration into \sal{} as well as the logging of browser events.\\
A frontend for participants & The \studyfrontend{} guides the participants through studies from beginning to end and can only proceed one step at a time. %
\\
Interaction logging & The system persists log data in the \backend. It supports logging native browser events, such as mouse, keyboard, and touch events. Furthermore, it allows the storage of custom data objects. Methods for data logging are provided in the \library{} and can be integrated by researchers as ``one-liners'', to enable easy data logging from inside prototypes. \\
Sharing of studies & \sal{} allows the import and export of studies. That way, study schemes can be shared easily as material for papers to support future research. \\
Keeping track of studies & In the long run, research teams can view the history of existing studies and can reuse and evolve their designs. \\
\bottomrule
\end{tblr}
\end{table}

\subsection{Motivation}
With our software system, we have the overarching goal to advance and substantiate the implementation of methods within web-based HCI research. The initial motivation to start the implementation stems from our experience: we re-implemented experiment logic and logging infrastructure several times across research projects. We found similarities when we compared our experience to related work in our domain. Researchers applied the same methods but gave no insights into implementing the methods. These researchers supposedly also spend time on the development of software and the integration of online services surrounding their research. We identified this as a challenge to provide a foundation for implementing methods in web-based studies. This allows researchers to focus on designing, developing, and studying interactions. Moreover, we support the vision of allowing researchers to share their study design and enable replication.

\subsection{Conceptual components}

From a conceptual view, \sal{} is composed of software components that closely mirror the conceptual components of experimental methodology. We describe these components and their features and properties in the following.

\begin{figure*}[t]
    \begin{center}
    \includegraphics[width=0.85\textwidth]{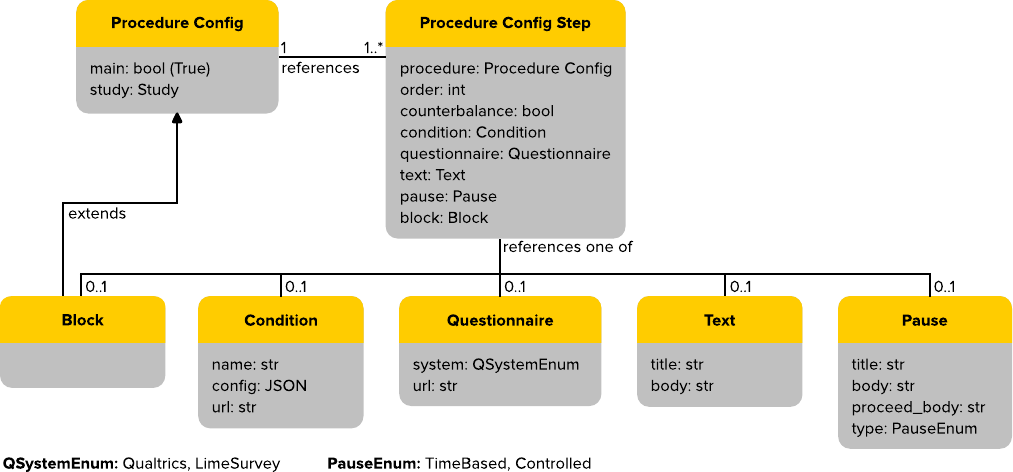}
    \caption{A class diagram of the Procedure Config used in the \backend{} to generate procedures. When a researcher sets up a procedure through the \controlpanel{}, a Procedure Config is created. This Procedure Config has multiple Procedure Config Steps. Each Procedure Config Step has an order number and a flag for counterbalancing. Furthermore, it is assigned a concrete procedure element, such as a Condition, Questionnaire, Text, Pause, or Block. A Block extends Procedure Config and can have multiple Procedure Config Steps. This way, a block can combine multiple elements into one element.}
    \Description{This figure shows a class diagram that is used in the \backend{} for setting up procedures. There are two rows. The top row shows two classes: procedure config and procedure config step. The bottom row shows five classes: block, condition, questionnaire, text, and pause. Block has an arrow to procedure config since it is a child of this class. Procedure config step has attributes to refer to a procedure config, and to one of the classes shown in the bottom row.}
    \label{fig:class-diagram}
    \end{center}
\end{figure*}

\subsubsection{Studies}
A study in our system is the main entity of an experiment and summarizes all corresponding entities (e.g. conditions, questionnaires, log data). A study can be closed access or open access, i.e. private or public. It has a start and end date. Furthermore, researchers can add a description and consent information, such as a privacy policy. For a closed-access study, researchers have to provide a distinct access link to the participants. For an open-access study, everybody who knows the link to the experiment can take part in the study.

\subsubsection{Conditions}
A condition is an aspect that researchers investigate by conducting an experiment, for example, a prototype that offers a novel UI or interaction method. In other words, each condition is typically one level of an independent variable. A condition in our system can be considered a proxy element that points to a deployed prototype on the web. It is possible to attach a config to a condition. With this config, a prototype can be configured automatically. Researchers only have to deploy one prototype, which reads the current config, for example, to alternate interface elements or interaction methods. In particular, quantitative data is collected with conditions.

\subsubsection{Questionnaires}
A questionnaire allows researchers to collect subjective data. Similar to a condition, a questionnaire can be considered a proxy element pointing to an external questionnaire service. Identifiers and study IDs are transferred to the survey tool through the questionnaires' GET parameters. After an experiment is finished, it is possible to collect data from the questionnaire service and match it with the \sal{} participant IDs.

\subsubsection{Text Pages}
A text page communicates information to participants within an experiment, such as a welcome text, task briefing, or general instructions. Text pages allow for HTML markup and embedding of videos. 

\subsubsection{Pauses}
A pause element allows for time-delayed experiments, for example, for interrupted time-series designs \cite{shadish_experimental_2002}. In such experiment designs, an intervention is caused during a procedure to measure its effects. In \sal{}, researchers potentially intervene in the experiment while a pause step is displayed, for example, by modifying a prototype. Researchers can define a concrete information page about the pause, which is displayed to the participants as a text page. A pause can be either time-based or manually controlled. With time-based pauses, the researcher can determine the pause length (e.g. three days). After the predetermined time has passed, participants will be allowed to proceed with the experiment automatically. With manually controlled pauses, researchers have full control over the pause of each participant. Only when an experimenter ends the pause of a participant will be able to proceed.

\subsubsection{Blocks}

A block is a logical element that allows researchers to combine several other procedure elements into one. In within-subject designs, the procedures might require not only counterbalancing the conditions but also keeping a task briefing and a questionnaire together with each condition. For example, researchers could combine a task briefing, condition, and questionnaire into one block, and then select the block for counterbalancing with other blocks. 

\subsubsection{Procedures}
A procedure represents the generalized experiment procedure. It is an ordered list of subsequent Conditions, Questionnaires, Text Pages, and Pauses. A procedure is defined by the experimenter, who selects which elements are counterbalanced by the system. The system internally uses a procedure config for setting up Procedures, as can be seen in \cref{fig:class-diagram}. Ultimately, this procedure config is then used to generate a set of procedures, for example, counterbalancing will result in multiple variants of the generalized experiment procedure. These counterbalanced variants of procedures are assigned to participants. Further information on typical procedures can be seen in \Cref{sec:use-cases}.

\subsubsection{Participants}
A participant in our system represents one person taking part in a study. Participants must retrieve an invite link for closed-access studies. A participant has a random UUID, i.e. Universal Unique Identifier, used to uniquely identify a participant and to log that person's interactions. Note that this does not require logging of personal data, the UUID is a random alphanumerical string. As long as researchers do not record identifying data, it is not trivial to reveal the real person taking part in the study. IP addresses are not recorded.

A participant's progress in the procedure is logged automatically. Persisting the progress can be useful, for example, to determine whether a participant dropped out of a study. 

\subsection{Navigator}
\begin{figure*}[t]
    \begin{center}
    \includegraphics[width=0.75\textwidth]{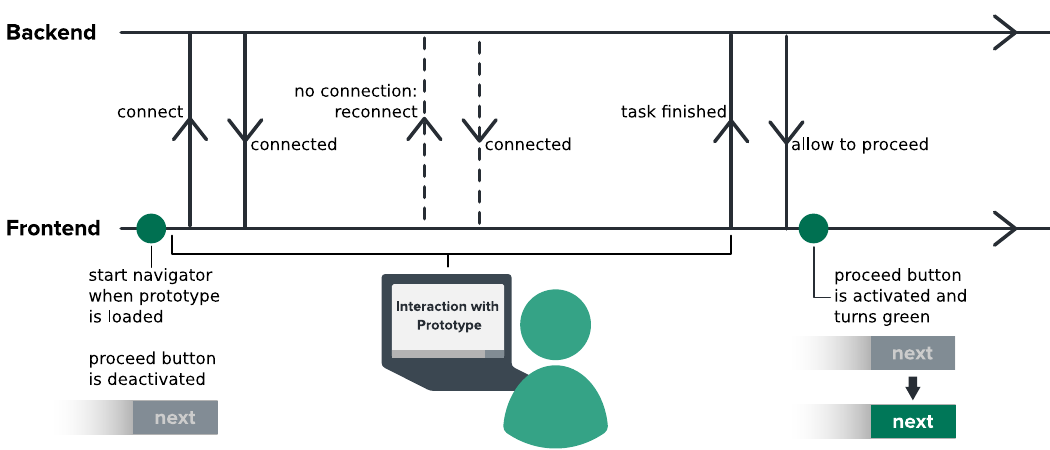}
    \caption{The Navigator is the conceptual component of \sal{} which controls the procedure. When a participant views a questionnaire or prototype, the Navigator decides whether the participant is allowed to proceed. The diagram is read from left to right and depicts the protocol. The Navigator connects at the beginning of a questionnaire or prototype, tries to reconnect if the connection fails, and waits for a signal when the task is finished for the participant to continue.}
    \Description{This figure visualizes the Navigator component that is used in the \studyfrontend{} to control the procedure flow. The figure shows two timelines: the top line visualizes the backend, the bottom line the frontend. Vertical lines visualize the communication between the backend and frontend. The communication reads from left to right: Frontend ``connects'' to the backend, and the backend confirms ``connected''. If the connection is lost, the frontend sends a ``reconnect'' to the backend, the backend confirms ``connected''. When the task is done, the frontend sends as ``task finished'' and the backend responds ``allow to proceed''. Annotations at the bottom show that this timeline captures the connection of a prototype to the backend. In the bottom left is a gray ``next'' button that is deactivated initially, then a participant interacts with the prototype. Finally, after the person is allowed to proceed, the ``next'' button turns green, as shown in the bottom right of the figure.}
    \label{fig:navigator}
    \end{center}
\end{figure*}

On a conceptual level, the Navigator is the part of \sal{} which controls the procedure of a participant. A participant is only allowed to proceed with a study, for example, if the task of a condition is fulfilled. When the participant starts a condition or questionnaire, the \studyfrontend{} connects to the Navigator and listens to the \backend{}. As soon as the participant is done with the task, a function call from the \sal{} \library{} from within the prototype updates the Navigator on the server. After that, the \backend{} sends a signal to the \studyfrontend{} to allow the participant to proceed with the study (e.g. enables a ``next'' button). The Navigator is implemented as Server-Sent Events for browser compatibility reasons. The flow diagram in \Cref{fig:navigator} shows an example of the Navigator sequence.

\subsection{System overview}
The architecture of \sal{} is split into four distinct software components: 1) \backend, 2) \controlpanel, 3) \studyfrontend, and 4) \library. Each component addresses certain requirements of researchers and participants. \Cref{fig:system-overview} visualizes how these components interact with each other. Overall, the \backend{} provides a REST API for communication. All data is stored in the central database of the \backend. The \controlpanel{} and \studyfrontend{} call requests on the \backend's API. The \library{} needs to be integrated into prototypes, for example, to log interaction data and make callbacks to the \backend. Questionnaires implement a callback to the \backend{} to update the procedure state. The research team configures and manages experiments using the \controlpanel{} within a web browser. Participants take part in studies via the \studyfrontend{} on their own devices within a web browser. 

\sal{} components that are implemented as web applications, such as the \controlpanel{} and the \studyfrontend{}, are served by a web server. The prototypes and questionnaires are hosted on the web as well. It is up to the research teams to decide where these components are hosted. It is possible to host them on one machine or split them, for example, for performance reasons. As a practical recommendation, we host the \backend, questionnaires, and \controlpanel{} on one machine. Prototypes run on other instances that allow for more performance, such as servers with access to GPUs to support computationally expensive operations, such as ML model inferences.

We designed \sal{} as a low-opinionated software system. Researchers can choose their preferred tech stack for implementing their web prototype and their preferred web survey tool for implementing questionnaires. We only require that the web prototypes support JavaScript for integrating our \library{} and that the survey tool passes through GET parameters for including a backlink in the end-of-survey message, to link back to \sal{}.

\subsection{Technical components}

The \sal{} system consists of four software components. These components are all implemented for web-based applications. However, it is possible to port those components that run on participants' devices to other platforms. The \library{} and the \studyfrontend{}, for example, could be ported to work in VR applications or on native mobile applications. Next, we describe our implementations of the technical components of \sal.

\subsubsection{\backend{}}

The \backend{} offers a REST API for communication and holds the complete experiment logic. It is implemented with Python 3.8. The \backend{} builds on FastAPI\footnote{\url{https://fastapi.tiangolo.com/}}, a high-performance web framework for building APIs. FastAPI is compatible with the OpenAPI\footnote{\url{https://github.com/OAI/OpenAPI-Specification}} and JSON Schema\footnote{\url{https://json-schema.org/}} standards, generates API documentation with SwaggerUI,\footnote{\url{https://swagger.io/tools/swagger-ui/}}, and offers data validation with pydantic.\footnote{\url{https://pydantic-docs.helpmanual.io/}} The data is stored in a PostgreSQL database.\footnote{\url{https://www.postgresql.org/}} PostgreSQL allows us to combine the benefits of indexed data structures and the flexibility of JSON fields, e.g. for data logging. We use SQLalchemy\footnote{\url{https://www.sqlalchemy.org/}} as an Object Relational Mapper and migrations are carried out with alembic.\footnote{\url{https://alembic.sqlalchemy.org/en/latest/}} 

Server-Sent Events\footnote{\url{https://developer.mozilla.org/en-US/docs/Web/API/Server-sent_events}} are used to push data from the server to the client to control the procedure of a participant. Administrative users are authenticated via JSON Web Tokens,\footnote{\url{https://jwt.io/}} while participants are authenticated via Logger API keys in the request headers.

In particular, the \backend{} processes data, for example, creating and updating data and counter-balancing procedures. Another core function of the \backend{} is to persist log data. At the moment, the system only supports logging text-based objects. Future updates of the \backend{}' logging capabilities will also allow storing more complex data, such as images.

\subsubsection{\library}

The \library{} drives the \studyfrontend{} and enables interaction logging in prototypes. It is implemented in TypeScript and standardizes the client-side API calls. 

For use in prototypes, it provides methods to connect to the \backend{} and log interactions. It allows logging single interactions or logging batches of interactions to minimize HTTP requests. Native Browser events, such as mouse, keyboard, and touch events, can be logged with minimal effort. For example, this snippet logs a mouse click event:

\label{code:logging}
\lstdefinestyle{sal}{
    commentstyle=\color{darkgray},
    keywordstyle=\color{coralpink},
    numberstyle=\tiny\color{darkelectricblue},
    basicstyle=\ttfamily\footnotesize,
    breakatwhitespace=false,         
    breaklines=true,                 
    captionpos=b,                    
    keepspaces=true,
    numbersep=5pt,                  
    showspaces=false,                
    showstringspaces=false,
    showtabs=false,                  
    tabsize=2
}
\lstset{style=sal}
\begin{lstlisting}
element.addEventListener('click', event => {
  lib.logMouseInteraction('click', event);
});
\end{lstlisting}

For batched transmissions of interaction logs, events are first logged to the browser's localstorage, and later sent to the \backend{} in batches of a specified size.

For each event, \sal{} persists the native browser event, the event's timestamp, study id, condition id, participant id, and a custom JSON object. For custom events, \sal{} does not log a native browser event, but any provided JSON object. More examples of interaction logging can be found in the documentation in our project repositories\footnote {\url{https://github.com/StudyAlign}}.

\subsubsection{\studyfrontend{}}

The \studyfrontend{} is the user interface that participants see when taking part in a study. It is implemented with React\footnote{\url{https://reactjs.org/}} and Redux\footnote{\url{https://redux.js.org/}} in ES6. The main function of the \studyfrontend{} is to display the procedure steps and guide participants through a study. The \library{} is implemented in the \studyfrontend{} for API communication with the \backend. 

\begin{figure}
    \begin{center}
    \includegraphics{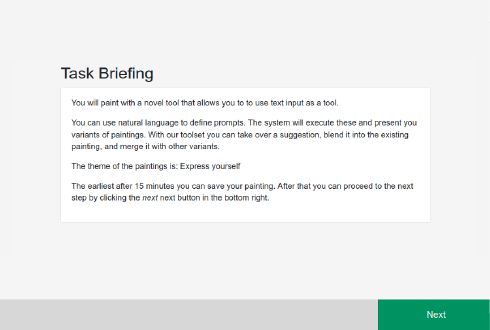}
    \caption{Screenshot depicting the actual UI of the \studyfrontend{}. In this screenshot, a task briefing that was implemented with a text page element is displayed. The text page displays a title and text instructions. The Navigator UI at the bottom of the screenshot allows participants to go ahead with the procedure by clicking on the ``next'' button in the bottom right.}
    \Description{This figure shows a screenshot of the UI of the \studyfrontend{}. It shows a text element of a study displaying a task briefing. It has a headline and a text body in the center. At the bottom of the screenshot is a toolbar showing a ``next button'' for proceeding with the experiment.}
    \label{fig:screenshot-task-briefing}
    \end{center}
\end{figure}

The UI of the \studyfrontend{} is minimalistic, see \Cref{fig:screenshot-task-briefing}: The \studyfrontend{} is separated into two areas. The larger area at the top displays text pages, prototypes, and questionnaires. The smaller area at the bottom of the UI displays a navigation bar to allow participants to proceed with the experiment via a ``next'' button. In conditions and questionnaires, the Navigator maintains a connection to the \backend{}, as described in \cref{fig:navigator}. The navigator controls the activation of this button. Activation calls from within the prototypes can request the frontend to activate the next button via the Navigator, for example, when a task has been completed.

\subsubsection{\controlpanel}

The \controlpanel{} is the user interface that researchers and experimenters use to configure and manage studies. It is implemented with React and Redux in ES6. The \controlpanel{} offers a dashboard, a wizard to configure studies and procedures, issuing participant IDs, and the import and export of studies as JSON data, as well as exporting interaction logs as CSV data. An overview of the features offered by the \controlpanel{} can be found in \cref{tab:controlpanel-features}.

\begin{table}[b!]
\caption{This table provides an overview of the features offered by the \controlpanel{}.}
\Description{Description}
\label{tab:controlpanel-features}
\begin{tblr}{lX}
\toprule
\textbf{Feature}           & \textbf{Description}   \\ \hline
Overview & The \controlpanel{} offers a dashboard as an overview of all studies. \\
Manage studies & Studies can be created and modified with a few clicks. \\
Share studies & Study schemas can be exported with only one click. Researchers can share their studies with colleagues and make them available as material for papers.\\
Study duplication & Studies can be duplicated. Researchers can adapt from existing studies to run subsequent studies with the same designs. \\
Log data preview \& export & Log data can be viewed from the browser, and also be exported for further analysis in R or Python. \\
\bottomrule
\end{tblr}
\end{table}

\begin{figure}
    \begin{center}
    \includegraphics{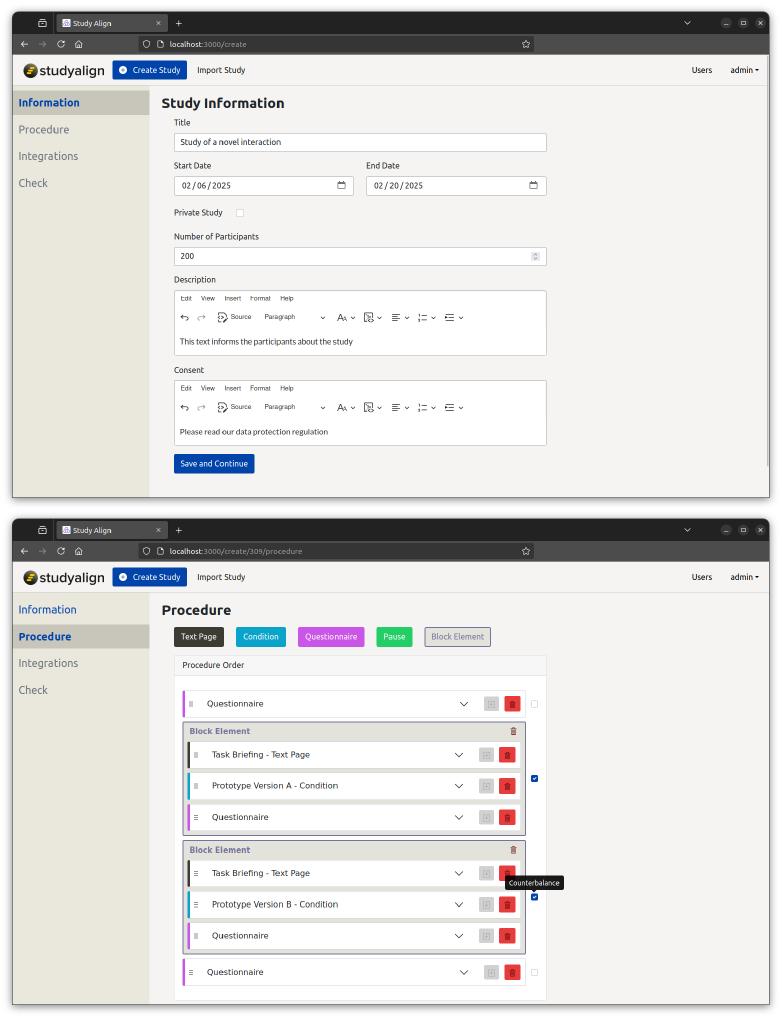}
    \caption{Two screenshots showing the first two steps of the \controlpanel{} UI for setting up experiments. The screenshot at the top shows the view for researchers to define general information about a study, such as the title, dates, number of participants, descriptive text, and consent text. The screenshot at the bottom shows the view where researchers can interactively compose a procedure. The buttons allow adding new steps which can be sorted through drag and drop. Nesting allows Block Elements to combine several other steps into one Block. Each step can be edited after unfolding the options. Elements that are selected through the checkboxes on the right will be counter-balanced by the system.}
    \Description{This figure shows two screenshots of the \controlpanel{} UI displaying the steps of setting up a study. Both screenshots have the same layout: A toolbar at the top with the \sal{} brand and a user menu. In the left a sidebar displaying the four steps: Study, Procedure, Integrations and Check. The screenshot at the top shows the step study. It is a form to define general information about a study, such as title, start and end date, number of participants, description text, and consent text. At the bottom of the screenshot is a button to get to the next step. The bottom screenshot shows the step procedure. It shows a view that researchers can use to create a procedure. At the top are buttons to add elements to the procedure list, such as text page, condition, questionnaire, pause, and block element. Below these buttons is the procedure list. Elements in this list can be sorted through drag and drop. Each element is collapsible and holds forms for defining the step. Nesting is displayed for block elements that hold multiple other steps. Icons on the right of each list element show a save button and a delete button. On the right of each list element are checkboxes for counterbalancing.}
    \label{fig:screenshot-procedure1}
    \end{center}
\end{figure}

\begin{figure}
    \begin{center}
    \includegraphics{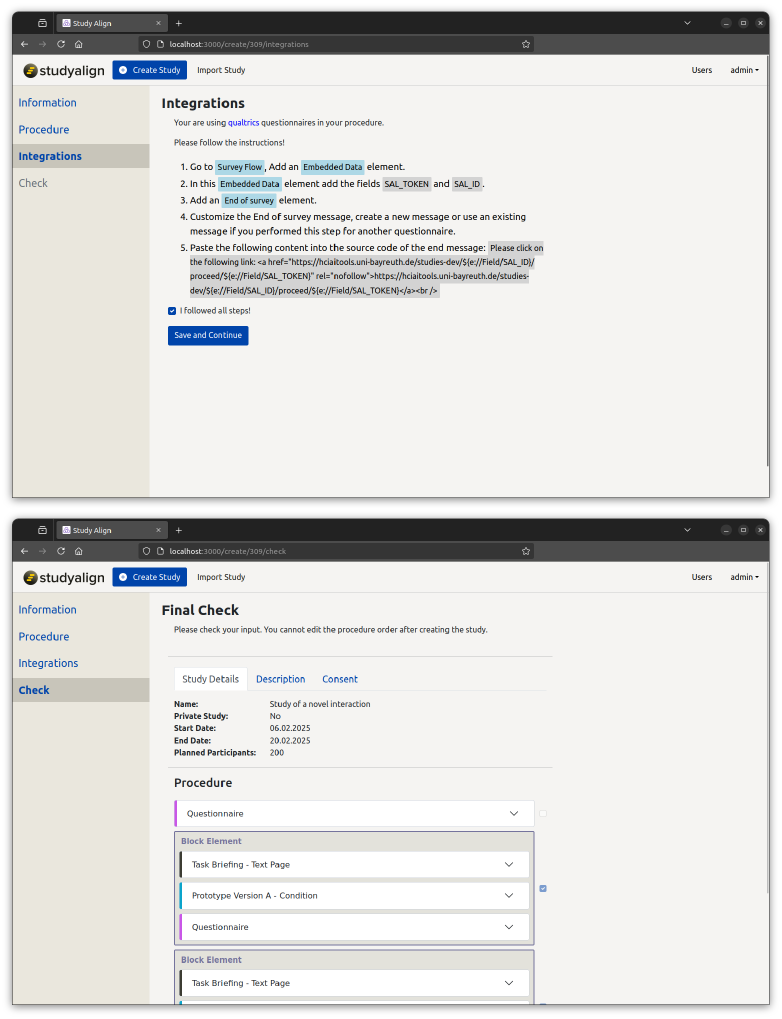}
    \caption{Two screenshots of the final two steps in the \controlpanel{} UI for setting up experiments. The screenshot at the top displays instructions for researchers on integrating questionnaires into the system. This step is only visible if researchers added a questionnaire to a procedure. The bottom screenshot shows a final overview of all input data so researchers can check if the experiment needs further editing.}
    \Description{This figure shows two screenshots of the \controlpanel{} UI displaying the steps of setting up a study. Both screenshots have the same layout: A toolbar at the top with the \sal{} brand and a user menu. In the left a sidebar displaying the four steps: Study, Procedure, Integrations and Check. The screenshot at the top shows the step integrations. It displays five steps for configuring questionnaires in external survey tools to work with \sal. Below these instructions is a checkbox that activates the button to proceed to the next setup step. The bottom screenshots show an overview displaying the provided information.}
    \label{fig:screenshot-procedure2}
    \end{center}
\end{figure}

In \cref{fig:screenshot-procedure1}, screenshots depict the first two steps in the UI when configuring an experiment. In these views, experiments are created step by step with a wizard-like interface. Researchers can easily configure each part of a study through forms, for example, by adding general information about the study, as can be seen in the screenshot at the top, and using drag and drop to modify the order of procedure steps, as shown in the bottom screenshot. \Cref{fig:screenshot-procedure2}, show the two final steps for setting up an experiment. The screenshot at the top shows instructions for integrating external survey tools, and the bottom screenshot is an overview of input data. Based on the input through this UI, the \backend{} will receive a formalized version of the procedure, similar to PlanOut \cite{bakshy_designing_2014}.

Log data can be exported from \sal{} in the \controlpanel{}. Researchers can then perform an analysis of this data with Python and R.

\subsection{External survey tools for creating questionnaires}

Our system supports any survey tool to integrate questionnaires into a procedure, as long as the tool supports passing URL parameters to the questionnaire for displaying backlinks. At the moment of writing this paper, we have gained experience with integrating Qualtrics\footnote{\url{https://www.qualtrics.com/}} and LimeSurvey\footnote{\url{https:/www.limesurvey.org/}} questionnaires into experiments. However, other popular survey tools such as SurveyMonkey\footnote{\url{https://www.surveymonkey.com/}} might support this approach as well: The \studyfrontend{} passes values through the URL to the questionnaire. The survey tool then stores these values to link questionnaire responses with the participants' UUID and puts the values into a backlink in the `'End of survey'' message, so that participants are allowed to proceed with the experiment. %

\section{Typical Study Designs}\label{sec:use-cases}

In this section, we present commonly used study designs in HCI and how they can be realized with \sal{} on a high level. In \Cref{sec:case-studies}, we present concrete case studies.

\begin{figure*}[t]
    \begin{center}
    \includegraphics[width=1\textwidth]{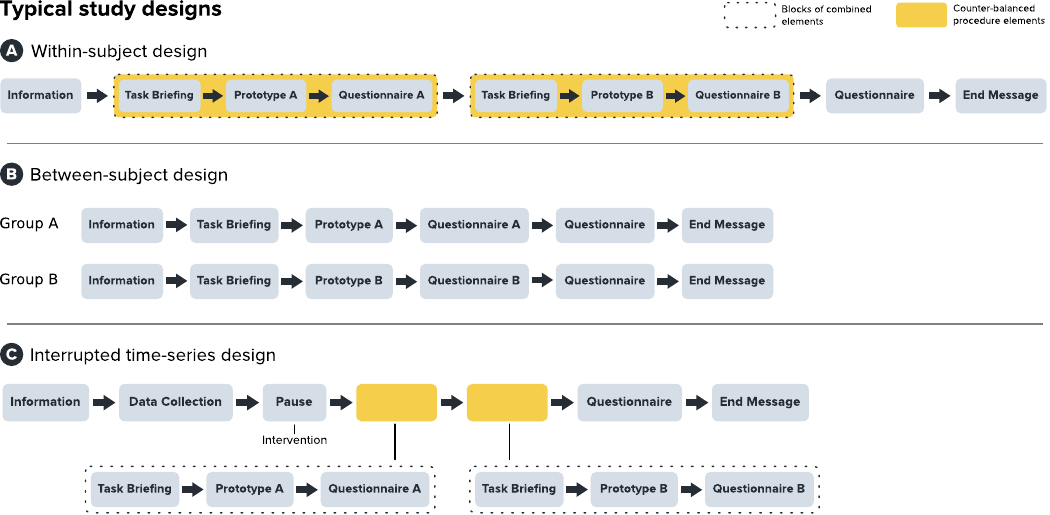}
    \caption{This figure shows typical study designs that can be implemented with \sal. The system supports within-subject, between-subject, and interrupted time-series designs. Example A shows a within-subject design: The task briefings, prototypes, and questionnaires in this example are combined into one ``block'' element. Such blocks of elements are then counterbalanced. Example B shows a between-subject design: A separate study is created for each group. It is possible to duplicate an existing study, which eases the creation of between-subject designs. Example C shows an interrupted time-series design: The depicted procedure is similar to a within-subject design but with an initial data collection step and an extended pause (e.g. multiple days), for example, to give researchers the time to create an intervention based on the collected data. This intervention then influences the prototypes in the remaining study procedure.}
    \Description{This figure shows typical study designs that can be implemented with \sal. The figure is structured into three rows, each showing one typical study design. Each design is visualized as a flow chart. The top row shows a within-subject design, which starts with an information block, followed by two block elements. Each block contains three blocks (task briefing, prototype, questionnaire). The block elements are highlighted to visualize counterbalancing. Followed by two blocks of visualizing a questionnaire and an end message. The row in the middle shows a between-subject design. It shows a flow chart for each group: Information, task briefing, prototype, questionnaire, questionnaire, and end message. The row at the bottom shows an interrupted time-series design. It begins with information, followed by data collection and a pause element that is annotated with ``intervention''. This step is followed by two block elements, each of which contains three steps (task briefing, prototype, questionnaire). The block elements are highlighted to visualize counterbalancing. These elements are followed by a questionnaire and an end message.}
    \label{fig:use-cases}
    \end{center}
\end{figure*}

\subsection{Within-subject designs}

In within-subject designs, the same conditions are presented to each participant, but the order of conditions is counterbalanced between participants. For this example, we assume an experiment to compare two user interfaces. At the beginning of the experiment, a briefing is displayed, and prior to each interface variant, a short task description is displayed. After each interface variant, a questionnaire is displayed, e.g. to measure the task load and usability. At the end of the experiment, another questionnaire is displayed to record socio-demographic data and to gather open-ended feedback. See \Cref{fig:use-cases} A for a visualization of the concrete procedure. Here, \sal{} counterbalances the order of blocks of prototypes and questionnaires between participants.

\subsection{Between-subject designs}

In between-subject designs, conditions are compared between groups. Only one condition is presented to people of one group. Typically, there are two or more groups of participants, depending on the number of conditions that should be compared. For this example, we also assume an experiment to compare two user interfaces, but between user groups. The procedure elements stay the same as in the example for within-subject designs, except only one of the user interfaces will be used and rated by each participant. See \Cref{fig:use-cases} B for a visualization of the concrete procedures. Between-subject designs can simply be realized in \sal{} by creating two studies each showing one of the prototypes.

\subsection{Studies with an interrupted time-series design}

In interrupted time-series designs, the effects of interventions are studied that occur in observations over time \cite{shadish_experimental_2002}. This design can be particularly useful for studies on interaction with AI, for example, when a model is fine-tuned or switched within an experiment.

For this example, we assume an experiment to compare the effects of personalized user interfaces on usability. This example uses a within-subject design with a data collection phase (usage of a baseline prototype) before comparing the two user interfaces. The pause in this example takes two or three days, enough time for the researchers to train an ML model that adapts a user interface according to prior collected user data. At the beginning of the experiment, a briefing is displayed, and before each interface variant, a short task description is displayed. Following each interface variant, a questionnaire assesses the perceived usability. A post-hoc questionnaire collects opinions on the personalized interfaces. After the initial data collection phase, a pause element informs the participant when the experiment continues. See \Cref{fig:use-cases} C for a visualization of this procedure. In this example, researchers end the pause for each participant individually and inform the participant via a messaging tool or email to continue. It is possible to configure time-based pauses that count down the time until a participant is allowed to proceed. However, we recommend using time-based pauses only for short windows of time to avoid drop-outs.

\section{Case Studies}\label{sec:case-studies}

\sal{} has been applied in more than ten studies for conducting experiments and collecting interaction data. Six of these studies have been published already \cite{dang_beyond_2022, dang_choice_2023, benharrak_writer-defined_2024, draxler_ai_2023, zindulka_exploring_2025, zindulka_content-driven_2025}. The other works are planned or in the process of being published, for example, \cite{lehmann_functional_2024}.

We use these studies as case studies in this section and provide specific details about how \sal{} enabled specific aspects of these studies. All of the presented case studies investigated text-based interactions in web prototypes.

\begin{figure*}[t]
    \begin{center}
    \includegraphics[width=1\textwidth]{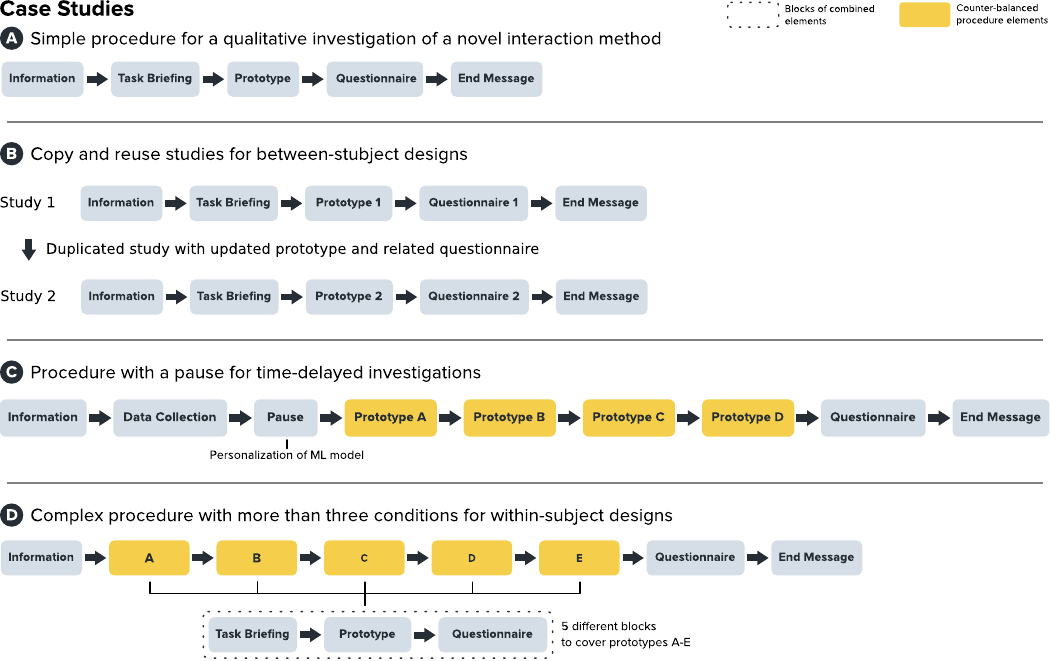}
    \caption{The procedures of the case studies implemented with \sal. These covered qualitative exploratory studies, qualitative and quantitative (mixed-methods) studies, one with experimenter interventions and a time delay (pause), and a comparative study.}
    \Description{This figure shows existing studies implemented with \sal{}. It is structured into four rows, each row showing one case study visualized through flow charts from left to right. The first row shows a simple procedure: information, task briefing, prototype, questionnaire, and end message. The second row visualizes the copy and reuse of a procedure. It shows two identical flow charts. Between these two charts is an annotation visualizing duplication. The procedures are assembled: information, task briefing, prototype, questionnaire, and end message. The third row shows a procedure for time-delayed investigations: information, data collection, pause (personalization of ml model), followed by four prototypes highlighted to visualize counterbalancing, after these steps follow a questionnaire, and an end message. The fourth row shows a complex procedure for five conditions in a within-subject design. It shows information, followed by five highlighted block elements to indicate counterbalancing. Each of these block elements points to another flow chart showing the steps, task briefing, prototype, and questionnaire to visualize five different blocks to cover the different prototypes. The final steps are a questionnaire and an end message.}
    \label{fig:case-studies}
    \end{center}
\end{figure*}

\subsection{Simple procedure for a qualitative investigation of a novel interaction method}

With \sal{}, an intelligent text editor was evaluated, which helps users to plan, structure, and reflect on the writing process by offering summaries of the text \cite{dang_beyond_2022}. The authors conducted a qualitative study to gain insights into users' interaction with the prototype. For their evaluation, the researchers implemented a procedure with four steps: 1) an intro to the study, explaining the prototype's features, 2) using the prototype, 3) a semi-structured interview, and 4) a final questionnaire. While taking part in the study, participants had a video call with the experimenter. 

All steps were implemented with our system in a procedure as shown in \Cref{fig:case-studies} A. The first steps were implemented with text elements that display study information, followed by the task briefing. In the third step, a condition element presented the prototype to the participants and collected interaction data. After completing the task with the prototype, a questionnaire element presented the questionnaire. The procedure concluded with an end message in a text element.
By using these elements, \sal{} streamlined this experiment setup and reduced the effort for interaction logging.

\subsection{Copy and reuse studies for between-subject designs}

This project \cite{lehmann_functional_2024} investigated two alternative versions of a text editor with AI features in two user studies. The initial prototype integrated AI in a conversational UI, while the alternative version integrated AI into a toolbar UI. %
The researchers implemented the procedure for each study with the following steps: 1) information about the study, 2) the task briefing, 3) prototype usage, and 4) a final questionnaire.

The two procedures of these experiments were implemented with \sal{} as shown in \cref{fig:case-studies} B: they displayed the study information on the initial page, followed by a text element that described the task, including an embedded video. The third step used a condition element to present the prototype and collect interaction data. After completing the task, a questionnaire element displayed a questionnaire. Finally, a text element was used to thank the participant for taking part in the study.

The second study with the iterated prototype was created by duplicating the first study with \sal. The only required changes were to replace the URL of the prototype and the questionnaire in the respective elements and to modify the task briefing to explain the new UI.

In summary, the ``duplicate'' feature of \sal{} made creating an alternative version for the follow-up study easy. The duplicated study could be modified conveniently through the \controlpanel.

\subsection{Procedure with a pause for time-delayed investigations}
Another study \cite{draxler_ai_2023} investigated the effects of personalized LLMs on authorship. This study required a break as part of its procedure -- it was paused for participants while the researchers adjusted the model for text generation. The researchers designed the procedure to start with information about the study, and then began with a questionnaire to collect data about each participant's writing. This initial data collection was followed by a pause of up to three days. In the meantime, the researchers used the responses on the questionnaire to prepare few-shot prompting for a text model. When this was completed, notifications were sent out to the participants to continue with the experiment. In the remaining study, participants used four different prototypes to write text supported by personalized and placebo-personalized ML models. In the final step of the procedure, participants had to fill out a questionnaire to provide feedback on the prototypes. Prolific was used to recruit participants.

This time-delayed experiment was implemented with \sal{} as shown in \cref{fig:case-studies} C: the initial information was displayed on the start page, followed by a questionnaire element that displayed a questionnaire. After that, a pause element was used to realize an experimenter-controlled pause: as soon as the researchers finished adjusting the generations of the model to match participants' writing styles, they allowed participants to go ahead with the experiment through a click in the \controlpanel{} and notified participants via Prolific. Four condition elements displayed four different AI prototypes. These conditions were counterbalanced using the \controlpanel{}. The final step was a questionnaire, implemented through a questionnaire element, followed by a text element with an end message that contained a backlink to Prolific.

In summary, \sal{} enabled this online study via Prolific while maintaining control of the fairly involved procedure, including participants' progression across multiple days. %

\subsection{Complex procedure with more than three conditions for within-subject designs}

A recent study \cite{dang_choice_2023} investigated how prompting influences writing behavior. The experiment started with information about the study, followed by an introduction to the task. In the third step, participants provided socio-demographic data. Afterwards, they used five interaction variants in the prototype, each followed by a questionnaire. A final step asked for general feedback on the interaction methods and thanked the participants.

This experiment was implemented with \sal{} as shown in \cref{fig:case-studies} D: the information about the study and GDPR consent was displayed on the initial page. In the first step of the procedure, a text element provided a written briefing. Then, a questionnaire element implemented the socio-demographic questionnaire. Afterwards, participants interacted with five prototypes, accompanied by a specific task briefing and a specific questionnaire. These groups were built with block elements. The resulting five block elements were counterbalanced by \sal{}. To collect general feedback, a questionnaire element followed at the end, and finally, a text element displayed an end message.

In summary, \sal{} lowered the effort of setting up and conducting this rather complicated within-subject design and its systematic counterbalancing, alongside offering interaction logging.

\section{Discussion}

We reflect broadly on our software system \sal{} and discuss benefits for research teams, and what we have learned from using it for conducting studies and close collaboration with other research teams. Furthermore, we highlight the opportunities that \sal{} provides to HCI research, considering replicating studies and research transparency.

\subsection{Benefits for research teams}

We expect that research teams benefit in many ways from using \sal{} as a standardized system for conducting web-based studies. On a high level, we see four aspects: 

\subsubsection{Keeping track of studies}
After running multiple studies with \sal{}, research groups will have a history of studies. This history enables group members to adapt from existing studies and serves as a knowledge base. The latter is even more important in the long run since group members switch regularly.

\subsubsection{Running subsequent studies}
Recent web-based studies ran multiple subsequent studies as part of an investigation \cite{jakesch_ai-mediated_2019, schmitt_characterchat_2021, todi_conversations_2021, zhang_method_2021, gero_sparks_2022, han_textlets_2020, buschek_impact_2021}. \sal{} eases the setup of such subsequent studies through its duplication feature and modification of studies with the \controlpanel. 

\subsubsection{Scaling of studies}
\sal{} allows for scaling studies while maintaining some control, for example, about the procedure. Researchers can modify the number of participants even in a running study. The \studyfrontend{} takes care of guiding participants through studies. 

\subsubsection{Robustness of methods}
As we laid out in our motivation for building \sal{}, we repeatedly re-implemented methods for our HCI studies. With such repeated implementations, it is likely that errors occur, and we had to test these implementations every time before conducting studies. \sal{} has proven its reliability in more than ten studies, providing a reusable and robust system for conducting web-based experiments.

\subsection{Lessons learned from conducting web-based experiments with \sal} %

\subsubsection{Proven interaction logging from more than ten conducted studies}
\sal{} has been used to run more than ten online studies. The interaction logging worked reliably in these studies, some of which have been published already \cite{dang_beyond_2022, dang_choice_2023, benharrak_writer-defined_2024, draxler_ai_2023, zindulka_content-driven_2025, zindulka_exploring_2025}. Furthermore, consider the case studies in this paper (\Cref{sec:case-studies}). To record interactions, we offer to log single interactions or to write them to the server in batches. The latter has the benefit of reducing HTTP requests and database transactions. We recommend relying on batched interaction logging for experiments that need to log a high number of interactions (e.g. keystrokes). The interaction logging is already implemented in the \library{} and can be reused across projects. Logging methods are accessed in prototypes through adding simple function calls, see \cref{code:logging} for an example. Thus, the \library{} effectively reduces programming effort for researchers.

\subsubsection{Configuration and management of studies need to be as easy as possible}

In the early beginning of developing \sal{}, we only made the API and a brief readme file available and offered no UI for administering studies. In particular, from our first study, we identified the need for a \controlpanel{} and a more detailed documentation. Research teams had to work with abstract methods and execute them manually to reach a specific study configuration. We improved this process with a written step-by-step guide, which informed the UI of the \controlpanel{} to set up and configure a study. With the \controlpanel{}, researchers can now create, configure, and manage studies comfortably from their browsers instead of writing JSON files and executing API requests manually.

\subsubsection{Run web-based experiments more efficiently}
Our initial motivation to create \sal{} was to enable researchers to set up and manage web-based research easily. By applying our system to concrete research projects, we demonstrated how it can support researchers in this regard and make setting up and running studies more convenient. 

Qualitative feedback from our collaborators showed that \sal{} improved efficiency, particularly for people executing studies (e.g. PhD students). One person, for example, gave us the feedback ``It is very convenient to have everything in one system. This helped me to carry out the study faster.''. In the same line, two other users responded ``I found the system and logging backend pretty cool, it made the work a lot easier'', and ``I think the biggest advantage of the system was that you have a unified user interface for questionnaires and prototypes. You can connect several parts on different websites and it still feels like one study. Also, I found it very good and easy to assemble the study from the text elements, the questionnaire, and the prototypes, which are then automatically counterbalanced.''. 

All collaborators asked for features, such as recruitment platform integration and the possibility of adding more participants while the study is running. By using a study link generated by \sal{}, people can be invited, for example, on platforms such as Prolific. We implemented the latter request: Allowing more participants to take part in the study is possible by increasing the number of participants in a study.

Furthermore, they asked for a UI for administrative purposes since they relied on the API for setting up the studies: ``At that time there was no GUI, so the handling was a bit difficult at first.''. As a result, the \controlpanel{} is now a core feature of \sal.

Since we involved external users working with the toolkit for collecting feedback, we cover the toolkit evaluation strategy \textit{usage} as identified by 
\citet{ledo_evaluation_2018}. We also cover their evaluation strategy \textit{demonstration} since \sal{} has been used to run more than ten studies and thus has proven its feasibility.

\subsection{More platforms than web browsers and web-based field experiments: VR, AR, native mobile apps, and lab studies}

\sal{} is based on web technologies, and the types of HCI experiments we built our system for mainly evaluate prototypes implemented as web applications (e.g. \cite{gero_sparks_2022, lee_coauthor_2022, yuan_wordcraft_2022}). While our case studies (\ref{sec:case-studies}) showed examples of studies on interaction with text models, it is also possible to work with other media in such prototypes, for example, images, videos, and games. Modern web browsers are powerful and flexible: they support these media types, providing great opportunities for prototyping and experimenting. 

However, \sal{} is not limited to web browsers. The clear separation of the \backend{} and the \studyfrontend{} allows researchers to port our system to other platforms. To port \sal{} to another platform, researchers would only have to port the \library{} and \studyfrontend. To run experiments in virtual reality, for example, the \library{} could be ported to C\# and the \studyfrontend{} to C\# in Unity. For AR and native mobile apps, the \library{} and \studyfrontend{} could be ported to, for example, Android and iOS. In contrast, the \backend{} and \controlpanel{} are independent of this and thus do not require any changes.

The control and flexibility offered by \sal{} is also available for lab studies. We originally designed the system for web-based field experiments, yet it is possible to use it to run studies in the lab. To conduct a lab study with \sal{}, researchers host the software components on a local network or give access to already hosted instances on the web. The participants then take part in the study via the \studyfrontend{} as usual, but in the lab.

\subsection{Transparency by sharing prototype code and study materials as a best practice}

\sal{} facilitates research transparency in two ways: First, the \controlpanel{} enables multiple researchers to collaborate on one experiment. For instance, interdisciplinary teams benefit from this when coordinating joint projects. Second, \sal{} allows for replicating experiments and sharing study designs in a formal way -- as structured JSON files. This feature contributes to similar efforts from close fields of research, such as cognitive sciences \cite{almaatouq_empirica_2021, lange_just_2015, henninger_felix_labjs_2022, hartshorne_thousand_2019}, and enables researchers to share study designs in a formalized way to ease replication.

More broadly, we would like to encourage researchers to share other material, such as prototypes and datasets. Recent work showed that researchers still hesitate to publish their prototypes since they consider them unfinished and see commercial value in their applications \cite{wacharamanotham_transparency_2020}. However, we argue that in a field of research that creates prototypes -- and analyzes interactions through these -- the ``default'' expectation should be to disclose such methods and artifacts as well. With shared materials, other researchers will be able to review methods more thoroughly. This change in culture would allow, for instance, building on already existing prototypes and replicating study designs. 

With \sal{}, we contribute a first effort in this direction by offering the function to share not only the study design as a file but also to export a complete study, for example, including logged data. Researchers only have to add their prototypes and questionnaires to publish complete material.

\subsection{The future of \sal}

We make \sal{} publicly available as an open-source project and implement it within research projects and collaborations with other researchers. While our personal use motivates us to maintain the project, we expect the project to grow. Typically, extensions add more workload to maintain a codebase. To distribute this workload, we plan to encourage the community to take part in the design decisions and development of the system, for example, by offering tutorials. %

On the technical side, we designed the software architecture to be maintainable and modular, and consider it to be well-structured. This makes our system flexible and extensible. For future releases, we plan, for example, to integrate more questionnaire services and recruitment platforms, and to extend the logging capabilities. Furthermore, high-load tests of the logging API are required for large-scale user studies with thousands of parallel participants.

\section{Conclusion}

We introduce \sal, a software system for conducting web-based research, such as online studies and web-based experiments. Our software system consists of several building blocks: 1) a \backend{} that ties together study designs and offers a central store for interaction logging, 2) a \controlpanel{} for administrative purposes that can be used to configure and organize studies conveniently, 3) a \studyfrontend{} that streamlines procedures and guides participants through experiments step by step, and 4) a JavaScript \library{} to integrate \sal{} into prototypes, for example, to connect to the \backend{} API and log interactions.

We derived the architecture of our system from the methodological decisions of related work and our own experience conducting online studies. The system offers control over web-based experiments in general and makes research more efficient since the experiment logic does not need to be re-implemented. The system has demonstrated usefulness in more than ten HCI research projects.

\section{Open source}

We publish \sal{} as an open-source project to the community. We hope to advance research on novel user interfaces and interaction methods by streamlining web-based experiments in HCI. The project can be found online: \url{https://studyalign.com}

\begin{acks}
We thank Jannek Sekowski and Niklas Markert for their outstanding contributions to the development of the control panel. This project is partly funded by the Bavarian State Ministry of Science and the Arts and coordinated by the Bavarian Research Institute for Digital Transformation (bidt). Funded by the Deutsche Forschungsgemeinschaft (DFG, German Research Foundation) –- 525037874.
\end{acks}

\bibliographystyle{ACM-Reference-Format}
\bibliography{main}

\end{document}